\documentclass[12pt]{article}
\usepackage{amsmath}
\usepackage{amsfonts}
\usepackage{amssymb}
\usepackage{graphicx}

\begin{document}
\setcounter{page}{0} \topmargin 0pt
\renewcommand{\thefootnote}{\arabic{footnote}}
\newpage
\setcounter{page}{0}

\begin{titlepage}
\begin{flushright}
NSF-KITP-06-84
\end{flushright}

\vspace{0.5cm}
\begin{center}
{\Large {\bf ADE and SLE}}\\

\vspace{2cm}
{\large John Cardy$^{a,b}$} \\
\vspace{0.5cm} {\em $^{a}$Rudolf Peierls Centre for Theoretical
Physics, 1 Keble Road, Oxford OX1 3NP, UK}  \\\vspace{0.3cm} {\em
$^{b}$All Souls College, Oxford}

\vspace{2cm} October 2006

\end{center}

\vspace{1cm}

\begin{abstract}
\noindent We point out that the probability law of a single domain
wall separating clusters in ADE lattice models in a simply
connected domain is identical to that of corresponding chordal
curves in the lattice O$(n)$ and $Q$-state Potts models, for
suitable $n$ or $Q$. They are conjectured to be described in the
scaling limit by chordal SLE$_{\kappa}$ with $\kappa$ rational and
$>$2. However in a multiply-connected domain the laws can differ
from those for the corresponding O$(n)$ or Potts model. The
correspondence also sheds light on the scaling limit of multiple
curves.

\end{abstract}

\end{titlepage}

\newpage

\section{Introduction}
Chordal Schramm-Loewner evolution\cite{Sch} (SLE) is the unique
family of conformally invariant measures on curves connecting two
distinct boundary points of a simply connected domain. It is
commonly conjectured (and in some cases proved) to describe the
scaling limit of suitably defined curves in critical equilibrium
lattice models, for example the O$(n)$ and $Q$-state Potts models.
These models are originally defined as lattice spin models for
integer values of $n$ and $Q\geq2$. However in the random curve
representation of either model as a gas of non-intersecting closed
loops, each loop is weighted with a factor $n$ (resp.~$\sqrt Q$)
and so it gives a probability measure on loops for all
non-negative values of these parameters. These measures are
however non-local in the sense that they are not product measures
over regions with a finite number of degrees of freedom. To get
local measures, one needs to go back to the spin representations,
which make sense only for $n=1,2$ and $Q=2,3,4$ in the regions
$n\leq2$ and $Q\leq4$ where these models are conjectured to have a
non-trivial scaling limit.

The importance of a local probability measure arises in the
connection with conformal field theory (CFT), because it implies
that the corresponding CFT should satisfy reflection positivity
and therefore be representable in terms of operators acting on a
space of states with non-negative norm. In any 2d CFT, the scaling
operators may be arranged into representations of two commuting
Virasoro algebras, and the above requirement then implies that
these representations be unitary. This fact was used by Friedan,
Qiu and Shenker\cite{Fri} to classify all such possible unitary
CFTs with central charge $c<1$. This leads to a discrete series of
possible values for $c$
$$
c=1-\frac6{m(m+1)}
$$
with $m$ an integer $\geq3$, while the allowed values of the
scaling dimensions $(h,\bar h)$ (conformal weights) of the primary
operators are restricted to take values in the Kac table
$$
h_{r,s}=\frac{\big(r(m+1)-sm\big)^2-1}{4m(m+1)}
$$
where $1\leq r\leq m-1$, $1\leq s\leq m$.

One important feature of these CFTs is the closure of the operator
product expansion (OPE) on a finite number of primary operators or
their descendants, with scaling dimensions within the Kac table
above. This also happens in a wider class of models, the minimal
models, which are labeled by a coprime pair of positive integers
$(p,p')$, with the formulae above replaced by
\begin{equation}
\label{cpp}
c=1-\frac{6(p-p')^2}{pp'}
\end{equation}
and
$$
h_{r,s}=\frac{(rp-sp')^2-(p-p')^2}{4pp'}
$$
where $1\leq r\leq p'-1$, $1\leq s\leq p-1$.

However this result does not determine exactly which of these
values occur in a given CFT. It was pointed out\cite{Car} that
this is highly constrained by modular invariance of the partition
function on the torus. This condition was solved by Cappelli,
Itzykson and Zuber\cite{Cap}, who showed that the resulting CFTs
can be labelled by Coxeter diagrams\cite{Cox} of type A, D or E
with Coxeter number $p$. By generalizing the restricted
solid-on-solid models of Andrews, Baxter and Forrester\cite{And}
(which correspond to the A series) Pasquier\cite{Pas1,Pas2}
constructed lattice ADE models, with local weights defined by the
adjacency matrix of G, which he showed, on the basis of Coulomb
gas and other arguments, give the CFTs corresponding to $p'<p$.
Later, Kostov\cite{Kos} defined dilute ADE models which correspond
to the other case.

In this note we recall these arguments in the context of
identifying candidates for curves in lattice models with local
weights which should have SLE as their scaling limit. In the case
of a simply connected domain with suitable boundary conditions, we
argue that these curves have  weights identical with those of the
O$(n)$ or $Q$-state Potts models with the same value of $c$, and
therefore if the latter converge to SLE in the scaling limit, so
do the corresponding curves in the ADE models. For other
geometries, however, the weights are not in general the same, even
in the scaling limit. This observation relates to the apparent
non-uniqueness of attempted definitions of SLE in
multiply-connected domains, and of connection probabilities of
multiple SLEs. We also show that, in the case A$_m$, the law of
$N<m$ curves is the same as $N$ `ordinary' multiple SLEs (with the
same value of $\kappa$) conditioned not to meet. In that case the
resulting partition functions agree with those recently proposed
by Bauer, Bernard and Kytola\cite{Bau} and Dub\'edat\cite{Dub}.

Most of the ideas in this paper are not new, in fact some of them
date back 20 years. The main purpose to recall them in the context
of SLE for those currently working in this active field.

The outline of this paper is as follows. In Sec.~2 we define the
dilute and non-dilute models for which the heights take values on
the nodes of any connected graph $\cal G$. We recall the arguments
of Pasquier which show that the can be reformulated as a gas of
non-intersecting loops, where each loop is weighted by $\Lambda$,
an eigenvalue of the adjacency matrix of $\cal G$. For the local
weights in these models to be real and positive, $\Lambda$ must be
the largest eigenvalue. The requirement that this weight should be
strictly less than than 2 (corresponding to CFTs with $c<1$) then
restricts $\cal G$ to be one of the ADE Coxeter diagrams.
Inclusion of $\Lambda=2$ ($c=1$) allows $\cal G$ to be an extended
Coxeter diagram of type $\hat{\rm A}$, $\hat{\rm D}$, $\hat{\rm
E}$, or A$_\infty$, which is the discrete gaussian model. The
other non-unitary minimal models then correspond to taking
$\Lambda$ to be a non-maximal eigenvalue. In general these have
complex local weights, although the loop gas measure still makes
sense as long as $\Lambda\geq0$.

In a simply connected domain with homogeneous boundary conditions,
there are only closed loops, and Pasquier's arguments show that in
any of the ADE models these are weighted exactly as in the
corresponding O$(n)$ or $Q$-state Potts models, and therefore
should have the same scaling limit. This has recently been
conjectured to be the Conformal Loop Ensemble (CLE$_\kappa$)\cite{She}. The
above identification of ADE and the other models is in agreement
with the idea that, given a value of $\kappa$, the CLE is a unique
conformally invariant measure on nested closed loops in a simply
connected domain.

In Sec.~3 we identify the clusters whose boundaries are supposed
to be described by SLE in the scaling limit. In the dilute ADE
model, these are clusters of adjacent lattice sites with the same
height. In the non-dilute model they are boundaries of clusters in
a Fortuin-Kasteleyn-type representation. By trivially extending
Pasquier's arguments we show how, with suitable boundary
conditions, chordal curves separating clusters whose heights
correspond to adjacent nodes of $\cal G$ have the same weights as
corresponding curves in the O$(n)$ and Potts models. Then, in
Sec.~4 we discuss the additional features which arise for multiple
curves and in multiply-connected domains.

After this work was completed a paper\cite{Fen} by P.~Fendley
appeared which covers some of the same material from a slightly
different point of view. In addition this paper discusses new loop
models corresponding CFTs with central charge $c>1$ which would be
very interesting to understand from the SLE perspective.

\section{ADE lattice models and loop gases.}
In this section we recall the definition of the local height model
associated with a graph $\cal G$, and the arguments of
Pasquier\cite{Pas2}  transforming it to a loop gas.
\subsection{Dilute ADE models.} This case is actually slightly simpler to discuss,
so we present it first. The dilute models were first introduced by
Kostov\cite{Kos} (see also Nienhuis\cite{Nie}), and shown to be
part of an integrable family of models satisfying the Yang-Baxter
relations by Roche\cite{Roc} and by Warnaar, Nienhuis and
Seaton\cite{Sea}. Kostov's models were originally defined on a
square lattice, but it is more natural, and symmetrical, to define
them on a regular triangular lattice as follows.

At each site $j$ of the triangular lattice is a height $h_j$,
which takes values on the nodes of a fixed connected graph $\cal
G$. We denote the adjacency matrix of $\cal G$ by $\bf G$. There
is a constraint whereby the heights at neighboring sites of the
triangular lattice must either be equal, or adjacent on $\cal G$.
Assuming that $\cal G$ has no cycles of length $\leq3$, this
implies that at least two of the heights around any elementary
triangle must be equal. The weight for a given configuration is
the product of the weights over all the elementary triangles of
the lattice. These are given as follows: if all the heights are
equal, the weight is 1; if the heights are, for example,
$(a,b,b)$, the weight is $x(S_a/S_b)^{1/6}$, where $S_a$ is a
strictly positive function on the nodes of $\cal G$, to be made
explicit later.
\begin{figure}[h]
\label{triangle} \centering
\includegraphics[width=6cm]{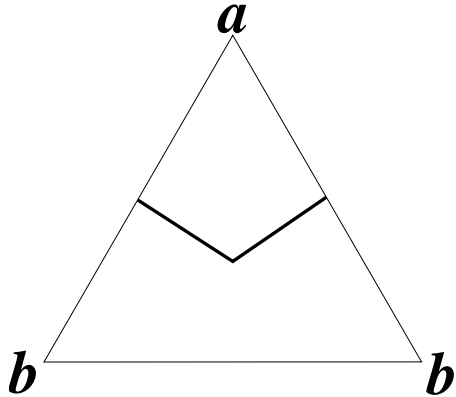}
\caption{An elementary triangle with unequal heights $(a,b,b)$ is
marked by a curve segment on the dual honeycomb lattice as shown.
This carries a weight $(S_a/S_b)^{1/6}$.}
\end{figure}

In the second case, one may mark the triangle as shown in
Fig.~\ref{triangle} by two straight lines connecting the midpoints
of the two edges joining the heights which differ to the center of
the triangle. These must join up with similar lines in the
neighboring triangles, and therefore must form closed,
non-intersecting, loops on the dual honeycomb lattice (or open
curves which end at a boundary.)  For the time being, restrict to
a simply connected domain where all the heights on the boundary
are fixed to take the same value. Then we have only closed loops.
Every loop separates a cluster of sites all with the same height,
say $a$, on its immediate interior from a cluster of sites with a
different height, say $b$, adjacent to $a$ on $\cal G$, on its
exterior. If one multiplies all the factors of $(S_a/S_b)^{\pm
1/6}$ around a given loop, this always gives $S_a/S_b$
irrespective of the shape of the loop.

The loops form a nested structure. Now sum over all allowed values
of the heights consistent with a given configuration of loops,
starting with the innermost clusters. Each sum has the form
$$
\sum_aG_{ab}(S_a/S_b)
$$
It is only straightforward to repeat this procedure at the next
level of nesting if this is some constant $\Lambda$ independent of
$b$: or, equivalently
$$
\sum_aG_{ab}S_a=\Lambda S_b
$$
That is, the $S_a$ should be the components of an eigenvector of
$\bf G$ with eigenvalue $\Lambda$. Taking this to be the largest
(Perron-Frobenius) eigenvalue then guarantees that all the $S_a$
are positive.

Summing iteratively over all heights of all the clusters then
gives a factor $\Lambda$ for each closed loop. This is to be
compared with the O$(n)$ model on the honeycomb lattice, which can
be written in terms of identical loops weighted with a factor $n$
for each loop, as well as a factor $x$.

This result shows that the law of the dilute ADE loops in a simply
connected domain with homogeneous fixed height boundary conditions
is identical to those of the $O(n)$ model with free boundary
conditions, at the same value of $x$, with the identification
$n=\Lambda$. If $n>2$, the O$(n)$ model is known either to be
non-critical or to exhibit only first-order critical behavior. The
condition that $\Lambda<2$ restricts $\cal G$ to be of the ADE
type (or ${\rm A}_{2m}/{\rm Z}_2$, which does not give a sensible
model.) $\Lambda=2$ corresponds to one of the extended ADE
diagrams or A$_\infty$. In each of the ADE cases,
$\Lambda=2\cos(\pi/h)$ where $h$ is the Coxeter number of the
graph.

The case A$_m$ is simple: we then have $\Lambda=2\cos(\pi/(m+1))$
and $S_a\propto\sin(\pi a/(m+1))$ with $a\in\{1,\ldots,m\}$. The
case $m=2$ corresponds to the Ising model on the triangular
lattice.

The O$(n)$ model for $n\leq2$ on the honeycomb lattice is
conjectured\cite{Nien2} to have a critical point at
$x=x_c=(2+\sqrt{2-n})^{-1/2}$. The scaling limit is also
conjectured, using Coulomb gas and CFT arguments, to correspond to
$\kappa\leq4$ where $n=-2\cos(4\pi/\kappa)$. Putting these
relations together in our case we have
$$
\kappa=\frac{4h}{h+1}
$$
where $h=m+1$ for the A$_m$ models.

The complete set of eigenvalues of $\bf G$ are given by
$\Lambda=2\cos(\pi h'/h)$ where $h'<h$ is one of the Coxeter
exponents of $\cal G$: for A$_m$ these take all integer values in
the range $1\leq h'\leq m$. The generalization of the above
relation is then
\begin{equation}
\label{lesser}
\kappa=\frac{4h}{h+h'}
\end{equation}
If we consider all possible A$_m$ models with $m\geq2$, this
includes all rational values of $\kappa$ in the interval $(2,4)$.
However only those in $(\frac83,4)$ have positive weights for the
loops.

The whole phase for $x>x_c$ is supposed be critical and
corresponds to $\kappa\geq4$, with the same relation between $n$
and $\kappa$. In that case we have, in the unitary case,
$$
\kappa=\frac{4h}{h-1}
$$
and more generally
\begin{equation}
\label{bigger}
 \kappa=\frac{4h}{h-h'}
\end{equation}
If we consider all possible A$_m$ models, this includes all
rational values of $\kappa>4$. However only those in $(4,8)$ have
positive weights for the loops.

\subsection{Non-dilute ADE models.}
These are most easily defined on a square lattice. Once again
there are heights at the vertices which take values on the nodes
of $\cal G$, but now heights at neighboring sites must be strictly
adjacent on $\cal G$ and are not allowed to be equal. If $\cal G$
has no cycles of length $\leq4$, this implies that at least one of
the diagonally opposite pairs of heights in each elementary square
must be equal. In any case, this constraint is enforced by the
weights (see below). Note that this also implies that the even
sublattice is populated by heights from the even nodes of $\cal
G$, and the odd sublattice by heights from the odd nodes (or vice
versa, as determined by the boundary conditions.) The weight for a
height configuration on the whole lattice is a product of the
weights around each elementary square.
\begin{figure}[h]
\label{square} \centering
\includegraphics[width=6cm]{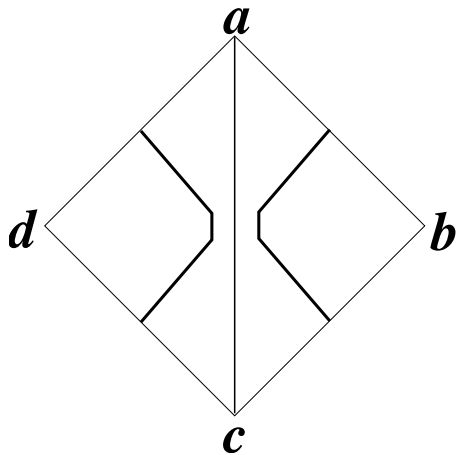}
\caption{Heights around an elementary square. If the first term in
(\ref{weight2}) is chosen, an edge is drawn connecting $a$ and $c$
as shown. This carries weight $(S_bS_d/S_a^2)^{1/4}$, which can be
distributed into factors $(S_b/S_a)^{1/4}$ and $(S_d/S_a)^{1/4}$
at the corners of the corresponding loop(s) on the medial
lattice.}
\end{figure}
Labelling the heights by $(a,b,c,d)$ (see Fig.~\ref{square}),
these are given by\footnote{This symmetrized version is equivalent
to the standard one in which the coefficient of the first term is
taken to be unity and the power in the second term is $\frac 12$.
This version is more suitable for transfer matrix and
Temperley-Lieb considerations.}
\begin{equation}
\label{weight2}
W(a,b,c,d)=\left(\frac{S_bS_d}{S_aS_c}\right)^{1/4}\,\delta_{ac}+
\left(\frac{S_aS_c}{S_bS_d}\right)^{1/4}\,\delta_{bd}
\end{equation}

The weight for a given configuration on the whole lattice may now
be expanded as a sum of $2^N$ terms, where $N$ is the total number
of elementary squares. In each elementary square we may draw an
edge connecting the sites with heights $a$ and $c$ if the first
term is chosen, or one connecting the sites with heights $b$ and
$d$ if the second term is chosen. The result is to decompose the
edges of the even sublattice into clusters. Within each cluster
all the heights are constrained to be equal. A similar
decomposition holds for the edges of the odd sublattice.
Neighboring clusters on the even sublattice are separated by
clusters on the odd sublattice and vice versa. The picture is
identical to that which occurs in the Fortuin-Kasteleyn (random
cluster) representation of the Potts model, with the clusters on
the even sublattice being identified as the clusters of Potts
spins, while those on the odd sublattice correspond to the
clusters of dual spins. As in the Potts case, neighboring clusters
can be separated by dense non-intersecting curves on the medial
lattice, which either form closed loops or end at a boundary.

The nice property of this picture is that the weights can now be
distributed at the corners of these curves (see
Fig.~\ref{square}). A corner which separates a cluster of height
$b$ from one of height $a$, such that the height within the acute
angle of the corner is $b$, carries a weight $(S_b/S_a)^{1/4}$.

Now consider a simply connected domain with homogeneous wired
boundary conditions. By wired we mean that the squares on the
boundary carry only the term in (\ref{weight2}) which corresponds
to there being an edge between the boundary sites. (Clearly this
requires that all the boundary sites be on the same sublattice,
say even.) We can then suppose that the heights on the boundary
are fixed to a given value. The edges connected to the boundary
form the boundary cluster. This cluster may be adjacent on its
interior to several clusters on the odd sublattice, which
themselves may be adjacent on the interior to clusters  on the
even sublattice, and so on. The curves on the medial sublattice
may only form closed loops. Each loop may touch only one cluster
internally, and one externally. The loops form a nested structure
as in the dilute case, separating clusters on which the heights
all agree. A loop enclosing a cluster of height $a$ and enclosed
by one of height $b$ carries a total weight $(S_a/S_b)$,
irrespective of the shape of the loop.

The summation over the heights for a given configuration of loops
then proceeds as in the dilute case. The weighting of the loops is
exactly as for the FK representation of the critical $Q$-state
Potts model, with identification $\Lambda=\sqrt Q$. The scaling
limit for $\Lambda\leq2$ is supposed to be described by
CLE$_\kappa$ as before. In this case we get only values of
$\kappa\geq4$, however, corresponding to the values in
(\ref{bigger}). Now the case ${\cal G}={\rm A}_3$ corresponds to
the Ising model. The heights on one sublattice are all fixed to
the same value $a=2$. On the other sublattice they take the values
$(1,3)$ corresponding to $\pm1$ in the usual formulation of the
Ising model. The $Q$-state Potts model in fact corresponds to
$\cal G$ being a star graph with a central node connected to $Q$
points. It is straightforward to check that the largest eigenvalue
of the adjacency matrix is $\Lambda=\sqrt Q$.

It is interesting to attempt to identify the boundaries of
connected clusters with the same value of the height variables in
the non-dilute ADE models (as for the Ising model.) For the $A_m$
models some progress can be made in that direction. Consider the
elementary squares around which the heights take the values
$(a,a+1,a,a-1)$. If we draw an edge connecting the two sites with
height $a$, this may be thought of as part of the boundary
separating clusters of heights $a\pm1$. It is straightforward to
see that only an even number of these edges can meet at any one
vertex of the square lattice, and therefore the set of such
cluster boundaries is the same as in the Ising case $m=3$.
However, to proceed further, one needs to sum over all possible
values of the heights within each closed loop, and this is
difficult is general. At the critical point, we expect to recover
weights for the loops which are the same as those in the dilute
O$(n)$ model on the square lattice (with some convention for
dealing with the vertices where 4 edges meet) but it may be that
this identification occurs only in the scaling limit. In the next
section we present some evidence for this conjecture.

\subsection{Models with $\kappa=4$.}
The models defined on graphs $\cal G$ whose adjacency matrix has
largest eigenvalue $\Lambda=2$ correspond to $\kappa=4$. In that
case $\cal G$ is either A$_\infty$ (the integers) or one of the
extended Coxeter diagrams, for example $\hat{\rm A}_m$. This has
$m$ nodes arranged in a ring. The first case corresponds to the
ordinary solid-on-solid model with the constraint that nearest
neighbor height differences take the values $(0,\pm1)$ (dilute
model) or strictly $(\pm1)$ (non-dilute model). The $\hat{\rm
A}_m$ models are particular realisations of the ${\mathbb Z}_m$
clock models, with same nearest-neighbor constraints. In these
cases the largest eigenvalue of the adjacency matrix is always
$\Lambda=2$, corresponding to $S_a={\rm const.}$, and the
arguments of the previous sections show that the law of loops (and
chordal curves -- see below) in a simply connected domain is
identical with that of the O$(2)$ model (dilute) or the FK 4-state
Potts model (non-dilute). (The latter in fact corresponds exactly
to $\hat{\rm D}_3$.) However in the clock models it is necessary
that $m\geq4$, to avoid 3-cycles. All these models are supposed to
have the same scaling limit given by the gaussian free field.

\section{Inhomogeneous boundary conditions and chordal curves.}
In this section we identify curves in these models which are
candidates for chordal SLE in the scaling limit. We show that the
law of these curves is identical to those in the corresponding
O$(n)$ and Potts models.

The dilute ADE models are simpler to discuss. To identify a
chordal curve we consider a simply connected domain and impose one
type of fixed boundary condition, say $a$, on one segment of the
boundary (the left boundary), and another, say $b$, on its
complement (the right boundary). This is strictly possible only if
$a$ and $b$ are adjacent on $\cal G$, in which case, in the
construction above, there is exactly one lattice curve $\gamma$
connecting the points $z_1$, $z_2$ where the boundary conditions
change. The sites immediately to the left of $\gamma$ all have
height $a$; those immediately to its right have height $b$. If $a$
and $b$ are not adjacent on $\cal G$, there must be several
lattice sites separating the left and right parts of the boundary.
This interval will go to a point in the scaling limit, but the
sites in between will be the starting points of more than one
curve. This situation therefore cannot give a single chordal curve
-- the right hand boundary of the height cluster connected to the
left boundary is not identical with the left boundary of the
cluster connected to the right boundary.

Given now a chordal curve $\gamma$ connecting $z_1$ and $z_2$, as
well as sets of nested closed loops in the simply connected
domains (with homogeneous boundary conditions) to its left and
right, we may sum over all the heights as before, beginning with
the most deeply nested clusters. This will give rise, as before,
to a factor $\Lambda$ for each closed loop. The curve $\gamma$
itself will accumulate factors $(S_a/S_b)^{\pm 1/6}$ for each left
or right turn. However since the number of right minus the number
of left turns is the same for every configuration, the relative
weights are all the same.

We conclude that the law of a single chordal curve in the dilute
ADE models in a simply connected domain is the same as in the
O$(n)$ model, and is therefore conjectured, in the scaling limit,
to be given by SLE$_\kappa$ with $\kappa$ given by
(\ref{bigger},\ref{lesser}).

For the non-dilute ADE models in the FK representation, consider
boundary conditions wired to a fixed height $a$ on the left
boundary, and to height $b$ on the right boundary. Once again, the
right boundary of the cluster attached to the left boundary, and
the left boundary of the cluster attached to the right boundary,
are adjacent only if $a$ and $b$ are adjacent on $\cal G$. (Note
that this means that the sites on the left and right boundaries
must be on opposite sublattices.) This then defines a chordal
lattice curve $\gamma$ which separates these two clusters. The
summation over heights goes as before, and we conclude that the
law of $\gamma$ is identical to the law of the  curve in the
$Q$-state Potts model, for the appropriate $Q$, with wired
boundary conditions on the Potts spins on the left boundary and
free boundary conditions (i.e. wired on the dual spins) on the
right boundary, which touches the right boundary of the FK cluster
attached to the left boundary. This is supposed to be given by
SLE$_\kappa$ in the scaling limit, with $\kappa\geq4$.

In this case, there is supporting evidence for this identification
from CFT. In the A$_m$ models, the boundary changing operator
between a boundary wired to height $1$ and to height $a$ was
identified by Saleur and Bauer\cite{Sal} as a $\phi_{1,a}$ operator in the
Kac classification. Here $1$ is a height at either extremity of
the diagram.  According to boundary CFT, the operator content at a
change of boundary conditions from $a$ to $b$ is given by fusing
$\phi_{1,a}$ and $\phi_{1,b}$. According to the fusion rules of
the A theories, this contains a $\phi_{1,2}$ operator if and only
if $a$ and $b$ are adjacent on the diagram. As was shown by Bauer
and Bernard, and Friedrich and Werner, the existence of a
$\phi_{1,2}$ boundary condition changing operator is necessary for
this to be an end point of an SLE (with $\kappa\geq4$.)

A similar argument may be applied to the height cluster boundaries
in the non-dilute case. Saleur and Bauer\cite{Sal} also considered boundary
conditions where the heights are fixed to $r$ on the boundary
sites, and to $r+1$ on the sites adjacent to the boundary, and
argued that the  the boundary condition changing operator between
$(1,2)$ and general $(r,r+1)$ corresponds to $\phi_{r,1}$ in the
Kac classification. The fusion rules can then be used to show that
the leading boundary condition changing operator between boundary
conditions $a,a-1$ and $a,a+1$ is a Kac $\phi_{2,1}$ operator,
which is a necessary condition for there to be an SLE with
$\kappa<4$ starting at the point where the boundary condition
changes. A candidate for this curve is the cluster boundary
described in the previous section, with some convention for
dealing with the 4-valent vertices.

It was recently shown by Riva and Cardy\cite{Riv} and
Smirnov\cite{Smi} that parafermionic observables of fractional
spin $s$ may be defined on the curves in FK Potts models which
weight the indicator function that curve goes along a given edge
of the medial lattice with a phase $e^{is\theta}$, where $\theta$
is the winding angle. These observables are discrete holomorphic
if $\kappa=8/(s+1)$. (Smirnov has also given a similar
construction for the O$(n)$ model.) Their correlators should
converge to those of holomorphic parafermions in the corresponding
CFT, although the convergence has so far been proved\cite{Smi}
only for the Ising case. The above result then implies that
holomorphic parafermions of spin $s=1\pm(2h'/h)$ should also exist
in the ADE models.

\section{Non-simply connected domains and multiple SLEs}
\subsection{Non-simply connected domains}
\begin{figure}[h]
\label{annulus} \centering
\includegraphics[width=7cm]{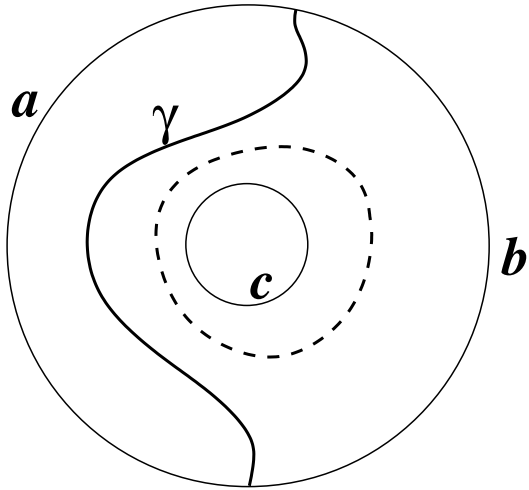}
\caption{A chordal curve in the annulus connecting two points on
the same boundary. The law of the this curve is sensitive to the
number of non-contractible loops, whose weights depend
non-trivially on the boundary heights.}
\end{figure}

In this section we show that the law of chordal curves in a
multiply-connected domain can in general be different for
different models with the same value of $\kappa$. As an example,
consider the annulus pictured as a simply connected domain with a
hole, and a chordal curve $\gamma$ which connects two point on the
same boundary of the annulus (see Fig.~\ref{annulus}). In the ADE
models, this can be enforced by imposing fixed height boundary
conditions, say $a$ and $b$, on the left and right segments of the
other boundary, and $c$ (not equal to $a$ or $b$) in the inner
boundary. In the Potts model, it would be necessary to condition
$\gamma$ not to hit the inner boundary. Without loss of generality
we may assume the $\gamma$ passes to the left of the hole, so that
the region to its right has the topology of an annulus.

In this case, as well as the chordal curve, there will in general
be non-contractible closed loops which wrap around the annulus, as
well as loops homotopic to a point. The summation over the heights
within the latter proceeds as before, giving a factor $\Lambda$
for each loop. However this is is not the case for the
non-contractible loops. In fact, if there are $N$ such loops, we
get instead
\begin{equation}
\label{ann}
\left[{\bf G}^N\right]_{bc}
\end{equation}
which may be interpreted as the number of $N$-step random walks on
$\cal G$ from $b$ to $c$. In general this depends on the choice of
$\cal G$ and on where $b$ and $c$ are located on $\cal G$. For the
Potts model, on the other hand, there is always simply a factor
$(\sqrt Q)^N$, essentially because all the other relevant
eigenvalues of $\cal G$ are zero.

This result implies that there is no unique way of prescribing the
driving term for SLE in an annulus or more complicated
multiply-connected domain (although we note that this difficulty
does not arise for curves which begin and end on different
boundaries of the annulus.) The distribution of the random
variable $N$ depends on the modulus of the annulus and can be
computed using Coulomb gas methods. It would be interesting to use
this to compute the SLE driving term for each of these models.

However we note that there is nevertheless a `canonical' version,
in which each non-contractible loop is counted with a weight
$\Lambda$. In the ADE models this may be realised by summing over
the boundary conditions $c$ on the inner boundary, and weighting
them with a factor $S_c$. This projects out the required
eigenvalue.

Also, as the size of the hole shrinks, keeping all other length
scales fixed, the typical value of $N$ goes to infinity. In that
case, only the largest eigenvalue $\Lambda$ of $\cal G$ is
important in (\ref{ann}), and the law of $\gamma$ becomes
universal once again. In this case it should be described by some
version of SLE$(\kappa,\rho)$.

\subsection{Multiple SLEs.}
The law of multiple chordal curves in a simply connected domain
has been considered by Bauer, Bernard and Kytola\cite{Bau} from
the point of view of CFT, and also by Dub\'edat\cite{Dub}. One of
the objects of interest is not the law of the curves themselves
(which is given by a version of SLE$(\kappa,\rho)$) but the
relative probabilities of the various ways the curves can connect,
or arch configuration, given a set of $2N$ points on the boundary.
These are given by ratios of partition functions in statistical
mechanics, which satisfy second-order partial differential
equations with respect to the positions of the points. These are
just the BPZ equations resulting from the fact that the CFT
operators which create the curves are $\phi_{1,2}$ (or
$\phi_{2,1}$) operators in the Kac classification (for $\kappa>4$
and $<4$ respectively.) However there is a question of which
boundary conditions correspond to a given arch configuration. A
clear answer to this is given by considering the equivalent
problem in the A$_m$ models.
\begin{figure}[h]
\label{arch}
\centering
\includegraphics[width=8cm]{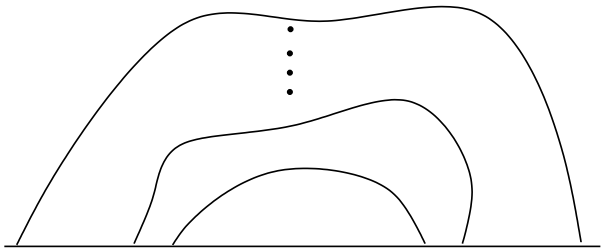}
\caption{An arch configuration in which $N$ curves are conditioned
to connect the boundary points a shown. Such a conditioning is
natural in the A$_m$ models if $N<m$.}
\end{figure}

Let us consider the geometry shown in Fig.~\ref{arch}. There are
$2N$ points on the real axis, connected by $N$ curves in the upper
half plane. Label these points in increasing order as
$(x_1,\ldots,x_N)$ and $(y_N,\ldots,y_1)$, and consider the
connection shown in Fig.~?? in which $x_j$ is connected to $y_j$.
Denote the partition function for this arch configuration by $Z_N(
x_1,\ldots,x_N;y_N\ldots,y_1)$. Its possible behavior as
neighboring points approach each other is given by the fusion
rules of CFT: $\phi_{1,2}\cdot\phi_{1,2}=\phi_{1,1}+\phi_{1,3}$,
which means that as $\delta=x_{j+1}-x_j\to0$, $Z_N$ can be
expressed as a linear combination of two terms, one of which
behaves as $\delta^{-2h_{1,2}}=\delta^{-(6-\kappa)/\kappa}$ and
the other as $\delta^{h_{1,3}-2h_{1,2}}=\delta^{2/\kappa}$.
Identical possibilities hold if $x_N$ approaches $y_N$.

Bauer, Bernard and Kytola identify an arch configuration in which
$x_1$ is connected with $y_1$ as being associated with a solution
where the dominant term as $x_1\to y_1$ is the fusion to the
identity operator $\phi_{1,1}$, while if neighboring curves are
conditioned not to meet, as for $x_j\to x_{j+1}$, it is associated
with the solution corresponding to fusion to the $\phi_{1,3}$
operator. In that case it is natural to normalize $Z_N$ so that,
as $x_N\to y_N$
$$
Z_N( x_1,\ldots,x_N;y_N\ldots,y_1)\sim
(y_N-x_N)^{-(6-\kappa)/\kappa}\, Z_{N-1}(
x_1,\ldots,x_{N-1};y_{N-1}\ldots,y_1)
$$
with $Z_0=1$ by convention.

Among all the solutions to the system of BPZ differential
equations with the above boundary conditions as $x_j\to y_j$ there
is in general one in which the operators $\phi_{1,2}(x_j)$ all
fuse to the $\phi_{1,N+1}$ operator, and similarly for the
$\phi_{1,2}(y_j)$. In that case it can be argued that in this
limit $Z_N$ has the form

\begin{equation}\label{ZN}
Z_N\propto \prod_{1\leq j<k\leq N}(x_k-x_j)^{2/\kappa}\,
\prod_{1\leq j<k\leq N}(y_j-y_k)^{2/\kappa}\,(y-x)^{-2h_{1,N+1}}
\end{equation}
where $x\sim x_j$, $y\sim y_j$.

By conformally mapping $y$ to infinity, Bauer et al\cite{Bau}
identify this boundary condition as selecting the partition
function for $N$ curves conditioned to go to infinity (rather than
forming arches among the points $x_j$). For small enough $N$ this
is completely consistent with the identification with curves in
the A$_m$ model. Suppose in the A$_m$ model we fix the boundary
conditions on each segment of the real axis to be
$(1,2,\ldots,N,N+1,N,N-1,\ldots,1)$ in order, where these integers
label nodes on the Coxeter diagram. Each of the points where the
boundary conditions change will be an end-point for a curve, and
will correspond to a $\phi_{1,2}$ operator. As long as $N\leq
m-1$, without any further conditioning these curves must connect
as in Fig.~\ref{arch}. Moreover if we now let $x_j\to x$ and
$y_j\to y$, we have boundary condition changing operators from $1$
to $N$ at $x$ and $y$, which are known\cite{Sal} to correspond to
$\phi_{1,N+1}$ operators.
\begin{figure}[h]
\label{connect} \centering
\includegraphics[width=8cm]{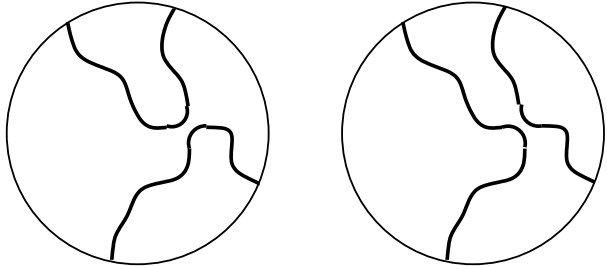}
\caption{The two different ways two curves can connect. The
relative probabilities depend not only on $\kappa$ but also on the
model and the boundary conditions.}
\end{figure}

However, these boundary conditions do not specify the relative
normalization of the partition functions for each arch
configuration, which are needed for the connection probabilities.
For example, in the case $N=2$, what are the relative
probabilities of the two arch configurations shown in
Fig.~\ref{connect}? The boundary conditions above fix each
partition function as a function of the cross-ratio $\eta$ of the
four points, up to overall constants. In the case of the Ising
model ($\kappa=3$) or of the boundaries of the FK Potts clusters,
where there is a ${\mathbb Z}_2$ symmetry (spin reversal or
duality) under the exchange of the two configurations, one may
argue\cite{Bau} that the relative weight must be 1 at the symmetry
point $\eta=\frac12$, which the determines the relative
probabilities for all values of $\eta$. However, for other ADE
models with the same value of $\kappa$, this ratio may be
modified. To see this it is useful to examine the case where the
two curves pass through the same square (in the non-dilute
models.) In that case the two different ways of connecting the
curves are weighted in the ratio
$(S_a/S_b)^{1/2}:(S_b/S_a)^{1/2}=S_a/S_b$. For the Ising or Potts
models this would be unity, but in general it is different. It is
also easy to see that the factor $S_a/S_b$ depends only on the
topology and not the fact that the curves pass through the same
square.

Once again this argument shows that certain aspects of these
random curves, even in the scaling limit, are model-dependent and
not solely determined by $\kappa$.

\noindent\em Acknowledgments. \em This work was carried out at the
Kavli Institute for Theoretical Physics, Santa Barbara, supported
in part by the National Science Foundation under Grant No.
PHY99-07949, in the course of the program on Stochastic Geometry
and Field Theory. I am grateful to several participants of that
program, including M.~Bauer, D.~Bernard, G.~Lawler, A.~Ludwig,
S.~Sheffield, S.~Smirnov and J.-B.~Zuber for useful suggestions
and critical comments, as well as to H.~Saleur.

\end{document}